# Local Solutions for Generic Multidimensional Resonant Wave Conversion

E. R. Tracy[a] and A. N. Kaufman[b]

[a]*Physics Department, William & Mary, Williamsburg, VA 23185-8795*
[b]*Lawrence Berkeley National Laboratory and Physics Department, UC Berkeley, Berkeley, CA 94720*

**Abstract.** In more than one spatial dimension, resonant linear conversion from one wave type to another can have a more complex geometry than the familiar 'avoided crossing' of one-dimensional problems. In previous work [1] we have shown that *helical* ray shapes are generic in a mathematical sense. Here we briefly describe how the local field structure can be computed.

## INTRODUCTORY COMMENTS

Resonant conversion between various wave types is exploited in RF heating schemes for fusion plasmas. Such conversion can occur in nonuniform plasmas where, for some spatial position $x_0$ and for a frequency $\omega_0$, two wave types 'a' and 'b' can have nearly equal wave vectors $k_a \sim k_b \sim k_0$. The resonance condition implies a matching of the local *phase velocities*, but still allows the two waves to have different *group velocities* and *polarizations*. Thus, the process cannot be reduced to one dimension, even locally, and the possibility of new physics arises. This is discussed more fully in [1-3].

In a separate paper [1], we have considered the question of what ray geometries might be 'generic' in multi-dimensions, and how the WKB connection coefficients can be calculated. We define the meaning of 'generic' conversion just after Eq. (7). This line of research is similar in spirit to the one-dimensional work of Littlejohn and Flynn [4]. In [1] we show that in systems with two or three spatial dimensions (implying that the ray phase space (**x**,**k**) is four- or six-dimensional, respectively), the ray geometry in conversion regions *cannot* be purely hyperbolic. Instead, it is generically a combination of hyperbolic motion in one two-dimensional subspace (analogous to a one-dimensional 'avoided crossing' or 'tunneling' region) and elliptical motion in another. Hence, the combined motion is helical. Such a combination of behaviors is, of course, not possible in the one-dimensional conversion problem. This result was independently derived using a different approach by Colin de Verdiere [5].

In [2] a tutorial introduction is given to ray-based analysis of multi-dimensional conversion, and in [1] we show that generic ray motion in multi-dimensional conversion is helical. Here we sketch the calculation of the wave field for helical conversion. Details will be provided in a longer paper.

## STATEMENT OF THE PROBLEM

Consider the linear wave equation (1) for the (three-component) electric field in a non-uniform, time-stationary, plasma:

$$\int d^2\mathbf{x}' dt' D_{jk}(\mathbf{x},\mathbf{x}',t-t')E_k(\mathbf{x}',t') = 0, \qquad j,k = 1,2,3. \tag{1}$$

Summation over repeated indices is implied. We assume that the plasma has two spatial dimensions for simplicity, but note that the approach can be generalized to higher dimensions. In addition, the WKB connection coefficients do not depend on the number of spatial dimensions, but simply on the fact that only two waves are undergoing conversion.

We assume that the wave equation is conservative and can be derived from an action principle. Using methods described in [2] we convert (1) into the form of a partial differential equation:

$$D_{jk}(\mathbf{x},-i\nabla,i\partial_t)E_k(\mathbf{x},t) = 0, \qquad j,k = 1,2,3. \tag{2}$$

Our goal is to solve (2) subject to some specified initial/boundary conditions. A standard tool for this analysis is the WKB method. WKB methods break down in conversion regions and must be augmented by a local treatment which provides an algorithm for calculating the WKB connection coefficients, as described below.

## THE LOCAL 2x2 WAVE EQUATION

In the conversion region, the electric field is expanded in the local form

$$\mathbf{E}(\mathbf{x},t) = e^{-i\omega_0 t} e^{i\mathbf{k}_0 \cdot (\mathbf{x}-\mathbf{x}_0)} \left[ \psi_\alpha(\mathbf{x})\hat{\mathbf{e}}_\alpha + \psi_\beta(\mathbf{x})\hat{\mathbf{e}}_\beta \right] \tag{3}$$

The *uncoupled* polarization vectors $\mathbf{e}_\alpha$ and $\mathbf{e}_\beta$ are (locally) constant and can be constructed using methods sketched in [2]. Inserting the ansatz (3) into (2) gives

$$D_{ij}(\mathbf{x}_0 + (\mathbf{x}-\mathbf{x}_0), \mathbf{k}_0 - i\nabla; \omega_0)\psi_j(\mathbf{x}) = 0, \qquad i,j = \alpha,\beta. \tag{4}$$

where

$$D_{ij} = \mathbf{e}_j^* \cdot \mathbf{D} \cdot \mathbf{e}_k. \tag{5}$$

(Here, and in what follows, the * notation refers to the Hermitian adjoint on vectors and operators.) Suppressing the $\omega_0$-dependence, and Taylor-expanding the wave operator about the conversion point, we have

$$\begin{pmatrix} \hat{D}_{\alpha\alpha} & \hat{D}_{\alpha\beta} \\ \hat{D}_{\alpha\beta}^* & \hat{D}_{\beta\beta} \end{pmatrix} \begin{pmatrix} \psi_\alpha(\mathbf{x}) \\ \psi_\beta(\mathbf{x}) \end{pmatrix} = 0 \tag{6}$$

where

$$\hat{D}_{ij} \equiv D_{ij}(\mathbf{x}_0,\mathbf{k}_0) + \nabla_x D_{ij} \cdot (\mathbf{x}-\mathbf{x}_0) - i\nabla_k D_{ij} \cdot \nabla. \tag{7}$$

Because the polarizations of (3) are the *uncoupled* ones, and because the conversion point lies on the dispersion surface for both uncoupled waves, the constant terms of the diagonal elements $D_{\alpha\alpha}$ and $D_{\beta\beta}$ are zero, while the off-diagonal term is typically a non-zero (complex) coupling constant $\eta$. In prior work [2,6], we assumed that the dominant terms in the vicinity of the conversion point were given by the first order corrections to the wave operator along the diagonal. This is, strictly speaking, only correct if the coupling constant is non-zero in the conversion *region* (not just at the conversion *point)*. Here we include the more general case as part of the analysis. Using an extension of methods described in [3] and more recent results from [1], it is possible to recast (5) into:

$$\begin{pmatrix} q_1 - i\gamma\partial_1 & q_2 + \Omega\partial_2 \\ q_2 - \Omega\partial_2 & q_1 + i\gamma\partial_1 \end{pmatrix} \begin{pmatrix} \psi_\alpha(q_1,q_2) \\ \psi_\beta(q_1,q_2) \end{pmatrix} = 0. \qquad (8)$$

Here $q_1$ and $q_2$ are new ray phase space coordinates that are linear combinations of the old **x** *and* **k**, and $\gamma$ and $\Omega$ are constants. Note that $\Omega$ is the rate of rotation about the conversion point of the elliptical part of the ray orbit, while $\gamma$ is the rate of exponentiation of the hyperbolic part [1]. Note also that the diagonal operators commute with the off-diagonal ones. We can now define a *generic* conversion to be one where all terms in the matrix operator of (8) are of equal importance.

## SOLUTION OF THE 2x2 WAVE EQUATION

Operating from the left with

$$\begin{pmatrix} q_1 + i\gamma\partial_1 & -(q_2 + \Omega\partial_2) \\ -(q_2 - \Omega\partial_2) & q_1 - i\gamma\partial_1 \end{pmatrix} \qquad (9)$$

and defining

$$\hat{D}_1 \equiv q_1 - i\gamma\partial_1, \quad \hat{D}_2 \equiv q_2 + \Omega\partial_2, \quad \hat{D}_3 \equiv q_1 + i\gamma\partial_1, \qquad (10)$$

leads to

$$\begin{pmatrix} \hat{D}_3\hat{D}_1 - \hat{D}_2\hat{D}_2^* & 0 \\ 0 & \hat{D}_1\hat{D}_3 - \hat{D}_2^*\hat{D}_2 \end{pmatrix} \begin{pmatrix} \psi_\alpha(q_1,q_2) \\ \psi_\beta(q_1,q_2) \end{pmatrix} = 0. \qquad (11)$$

Thus, the $\alpha$ and $\beta$ subspaces have decoupled. From the form of (11) it is seen that the $q_1$- and $q_2$-dependences separate. Further analysis shows that the $q_1$-dependence involves a parabolic cylinder-like equation (though not self-adjoint), and the $q_2$-dependence involves a self-adjoint equation like that of a quantum harmonic oscillator. The separation constant plays the role of an effective coupling constant (more precisely, the magnitude squared of the coupling constant). The general solution of (11) is a linear superposition of terms involving products of parabolic cylinder-type functions and harmonic oscillator eigenfunctions, with each term in the series having a different separation constant. The matching to incoming and outgoing WKB waves is done by first computing the expansion coefficients by fitting at large negative values of $q_1$ to the incoming WKB wave. Then, using the asymptotic behavior of the parabolic cylinder functions at large positive values of $q_1$, the outgoing WKB wave is calculated.

At the level of the ray picture [2], we find the following result: the incoming WKB wave is a *family* of rays, with an amplitude, phase and polarization assigned to each. The entire family of incoming rays follow helical orbits with helicity $\kappa=\Omega/\gamma$ as they pass through the conversion region and connect smoothly onto the family of outgoing *converted* rays. The *conversion coefficient* provides the amplitude and phase assigned to each of the outgoing converted rays. The *transmitted* family of rays also follows helical orbits. They each are paired with an incoming ray and assigned an amplitude and phase given by multiplication of the data on the incoming ray by the

*transmission coefficient*. Except for that small set of rays with effective coupling constant nearly zero (which generates an outgoing Gaussian beam) the transmission and conversion coefficients for each ray are identical to those obtained in [3] since they depend only upon the asymptotics of parabolic cylinder functions. This will be elaborated in a longer paper. We note that similar results were previously obtained by Littlejohn and Flynn [7].

## SUMMARY AND CONCLUSIONS

We have briefly described results recently obtained concerning resonant conversion of linear waves in multiple spatial dimensions. Our goal has been to understand *generic* behavior, rather than analyzing a particular physical model. The search for generic results, true for 'typical' representatives of a family of systems, can lead to very general results. However, we have found that magnetized plasmas are typically *not* generic in the sense we use here. This is because the gyro-orbits of magnetized particles exhibit symmetry around the local magnetic field. There are various ways in which genericity might be obtained; for example strong shear flows or fully three-dimensional gyro-orbits associated with complex magnetic field geometry might break the symmetry which makes magnetized plasmas nongeneric. This is work in progress.

## ACKNOWLEDGMENT

This work was supported by the USDOE Office of Fusion Energy Sciences.

## REFERENCES

1. E. R. Tracy and A. N. Kaufman, "Ray helicity: a geometric invariant for multi-dimensional resonant wave conversion", submitted to Physical Review Letters. Available online at http://arXiv.org/physics/0303086.
2. E. R. Tracy, A. N. Kaufman, and A. J. Brizard, Physics of Plasmas **10**, 2147-2154 (2003).
3. E. R. Tracy and A. N. Kaufman, Phys. Rev. E **48**, 2196-2211 (1993).
4. W. G. Flynn and R. G. Littlejohn, Ann. Phys. **234**, 334-403 (1994).
5. Y. Colin de Verdiere, "The level crossing problem in semi-classical analysis II: the Hermitian case", preprint, March 2003.
6. E. R. Tracy, A. N. Kaufman, and A. Jaun, Phys. Lett. A **290**, 309-316 (2001).
7. R. G. Littlejohn and G. Flynn, personal communication (unpublished).